\documentclass[final,5p,times, twocolumn]{elsarticle}

\usepackage{hyperref}
\usepackage{lineno}
\usepackage{comment}
\usepackage{amsmath}
\usepackage{float}
\usepackage{graphicx}
\usepackage{svg}
\usepackage{pdfpages} 
\modulolinenumbers[5]
\usepackage{pgfplots}
\usepackage{amssymb}
\usepackage{multirow}
\usepackage{ gensymb }
\usepackage{multicol}
\usepackage{makecell}
\usepackage{caption}
\usepackage{fixltx2e}

\usepackage{amsthm}
\usepackage{booktabs}

\makeatletter
\g@addto@macro{\UrlBreaks}{\UrlOrds}
\makeatother

\usepackage{subfig}
\usepackage{caption}
\journal{A}

\begin{document}
\begin{frontmatter}

\title{ \huge BAST: Binaural Audio Spectrogram Transformer for Binaural Sound Localization}

\author[2]{Sheng Kuang} 
\ead{sheng.kuang@maastrichtuniversity.nl}

\author[1]{Jie Shi} 
\ead{j.shi1@uu.nl}

\author[3]{Kiki van der Heijden}
\ead{kiki.vanderheijden@donders.ru.nl}

\author[1]{Siamak Mehrkanoon \corref{cor1}}
\ead{siamak.mehrkanoon@maastrichtuniversity.nl;s.mehrkanoon@uu.nl}

\cortext[cor1]{Corresponding author}




\address[1]{Department of Information and Computing Sciences, Utrecht University, Utrecht, The Netherlands}
\address[2]{Department of Data Science and Knowledge Engineering, Maastricht University, The Netherlands.}
\address[3]{Donders Institute for Brain Cognition and Behavior, Radboud University, Nijmegen, The Netherlands}


\begin{abstract}
Accurate sound localization in a reverberation environment is essential for human auditory perception. Recently, Convolutional Neural Networks (CNNs) have been utilized to model the binaural human auditory pathway. However, CNN shows barriers in capturing the global acoustic features. To address this issue, we propose a novel end-to-end Binaural Audio Spectrogram Transformer (BAST) model to predict the sound azimuth in both anechoic and reverberation environments. Two modes of implementation, i.e. BAST-SP and BAST-NSP corresponding to BAST model with shared and non-shared parameters respectively, are explored. Our model with subtraction interaural integration and hybrid loss achieves an angular distance of 1.29 degrees and a Mean Square Error of 1e-3 at all azimuths, significantly surpassing CNN based model. The exploratory analysis of the BAST's performance on the left-right hemifields and anechoic and reverberation environments shows its generalization ability as well as  the feasibility of binaural Transformers in sound localization. Furthermore, the analysis of the attention maps is provided to give additional insights on the interpretation of the localization process in a natural reverberant environment.
\end{abstract}

\end{frontmatter}

\section{Introduction}

Sound source localization is a fundamental ability in everyday life. Accurately and precisely localizing incoming auditory streams is required for auditory perception and social communication. In the past decades, the biological basis and neural mechanism of sound localization have been extensively explored \cite{batteau1967role,pickles2015auditory,van2019cortical,grothe2010mechanisms}. Normal hearing listeners extract horizontal acoustic cues by mainly relying on the interaural level differences (ILD) and interaural time differences (ITD) of the auditory input. These cues are encoded through the human auditory subcortical pathway, in which the auditory structures in the brainstem integrate and convey the binaural signals from cochleas to the auditory cortex \cite{pickles2015auditory,grothe2010mechanisms}. However, sound localization is frequently affected by the noise and reverberations in the complex real-word environment, which distort the spatial cues of the sound source of interest \cite{blauert1997psychophysics}. Yet, it is still not clear how the spatial position of acoustic signals in complex listening environments is extracted by the human brain.

Recently, Deep Learning (DL) \cite{lecun2015deep} has been proposed to model auditory processing and has achieved great success. These approaches enable optimization of auditory models for real-life auditory environment \cite{zhang2017deep,lee2021deep,gong2021ast,truong2021right}. In the early attempts, DL methods were combined with conventional feature engineering to deal with the noise and reverberation problems \cite{zhang2017deep,perotin2018crnn}. For instance, in \cite{zhang2017deep}, binaural spectral and spatial features were separately extracted, providing complementary information for a two-layer Deep Neural Network (DNN). Similarly, in \cite{lee2021deep,yoshioka2015far,park2017multiresolution}, deep neural networks were used to de-noise and de-reverberate complex sound stimuli. In a CNN-based azimuth estimation approach, researchers utilized a Cascade of Asymmetric Resonators with Fast-Acting Compression to analyze sound signals and used onsite-generated correlograms to eliminate the echo interference \cite{xu2019binaural}. However, most of these approaches highly depend on feature selection. To reduce this constraint, end-to-end Deep Residual Network (DRN) was recommended \cite{he2016deep,yalta2017sound}. Instead of selecting features from acoustic signal, raw spectrograms of sound was utilized in Deep Residual Network for azimuth prediction \cite{yalta2017sound}. DRN was shown to be robust even in the presence of unknown noise interference at the low signal-to-noise ratio. Subsequently, \cite{gong2021ast} proposed a pure attention-based Audio Spectrogram Transformer (AST) and achieved the state-of-the-art results for audio classification on multiple datasets. Although these DL-based methods have yielded promising results, however due to a lack of similar architectures to the human binaural auditory pathway, they may not resemble the neural processing of sound localization.  


To encode the neural mechanisms underlying sound localization, the performance of deep learning methods is commonly compared to the human sound localization behavior \cite{van2022goal,van2020modelling,francl2022deep}. For instance, \cite{francl2022deep} systematically explored the performance of binaural sound clips localization of a CNN in a real-life listening environment, however, its architecture does not resemble the structure of human auditory pathway. This issue has been addressed by utilizing a hierarchical neurobiological-inspired CNN (NI-CNN) to model the binaural characteristics of human spatial hearing \cite{van2022goal}. This unique hierarchical design, models the binaural signal integration process and is shown to have brain-like latent feature representations. However, NI-CNN \cite{van2022goal} is not an end-to-end model as it leverages a cochlear method to generate auditory nerve representations as model input. Furthermore, considering the wide range of frequencies of sound input, the convolution operations in NI-CNN mainly extract local-scale features and therefore may exhibit limitations for extracting global features in the acoustic time-frequency spectrogram.

In this study, we build on the success and barriers of previously proposed deep neural networks at localizing sound sources to further develop an end-to-end transformer based model for human sound localization, which captures the global acoustic features from auditory spectrograms. We aimed at (i) investigating the performance of a pure Transformer-based hierarchical binaural neural network for addressing human real-life sound localization;
(ii) exploring the effect of various loss functions and binaural integration methods on the localization acuity at different azimuths; (iii) visualizing
the attention flow of the proposed model to demonstrate the localization process.

\begin{figure*}[!t]
    \centering
    \includegraphics[scale=0.35]{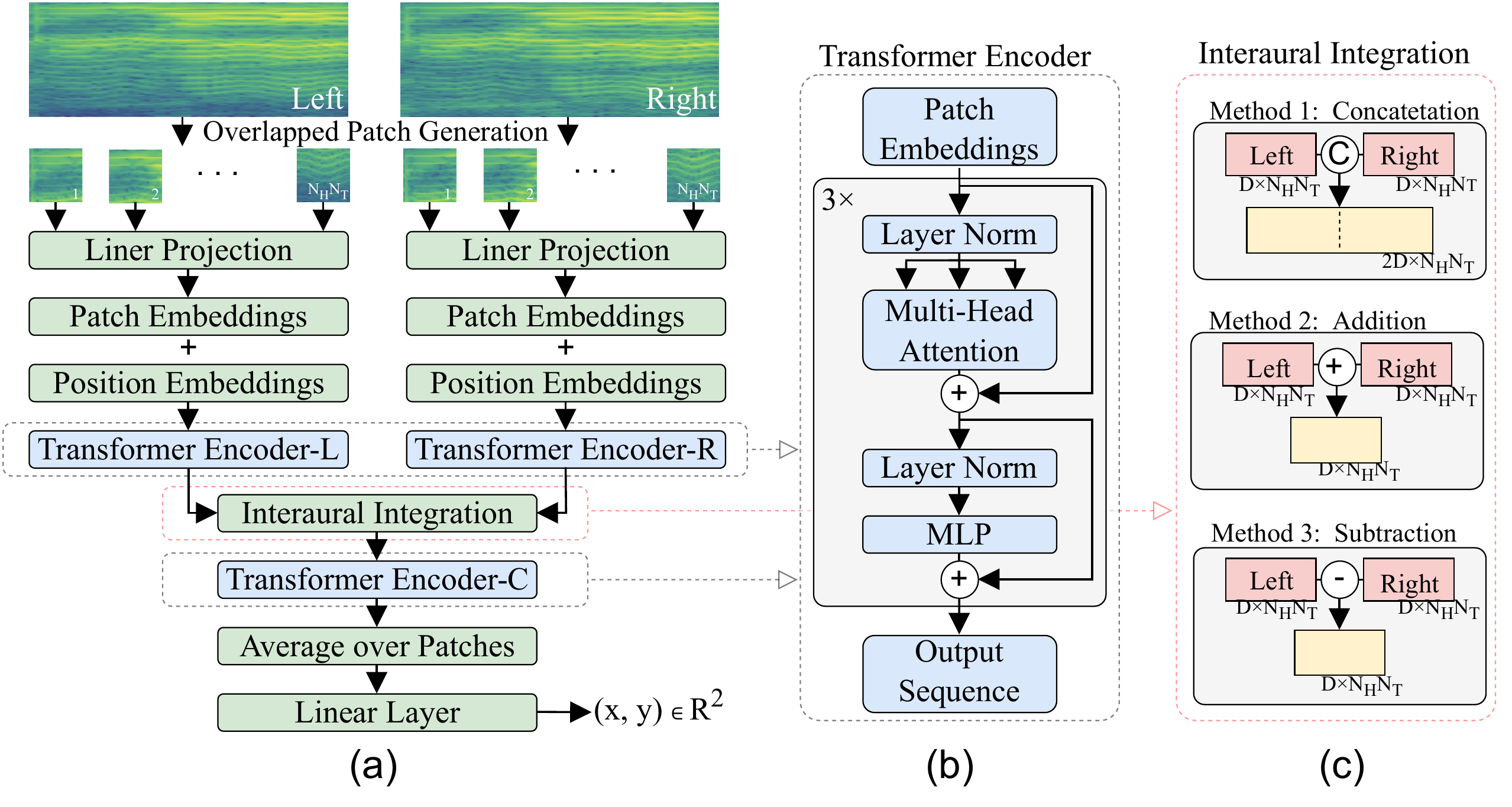}
    \caption{Architecture of the proposed Binaural Audio Spectrogram Transformer (BAST). (a) The architecture of the proposed model. (Here there are $N_{H}N_{T}$ number of patches). (b) The architecture of a single Transformer encoder. (c) Three examined interaural integration methods: concatenation, addition and subtraction.}
    \label{fig:archi}
\end{figure*}

\section{Related Works}

Binaural auditory models utilize head-related transfer functions to apply characteristics of human binaural hearing to monaural sound clips in order to simulate human spatial hearing. Conventional methods for sound source localization (SSL) have been based on microphone arrays and can be categorized into controllable beamforming, high-resolution spectrogram estimation, and time difference of sound techniques \cite{mathews2018multiple}. These conventional signal processing techniques are often used as baselines or input feature extraction for DL-based SSL methods. STEF (Short-Time Fourier Transform) \cite{durak2003short} approach is used to convert the time-domain signals from each microphone into the time-frequency domain. The STFT provides a representation of the signal in both time and frequency, allowing for the analysis of how the frequency content of the signal evolves. Mixture Model (GMM), commonly used in machine learning-based studies, calculates the probability distribution of the source location in reverberant environments \cite{ma2018robust}. Gaussian mixture regression (GMR) was extended later to localize multi-source sounds \cite{grumiaux2022survey}. Subsequently, model-based methods have been encouraged to extract ILD and ITD cues for DNN training \cite{zhang2017deep}. Compressive sensing and sparse recovery techniques are extensively applied in acoustics. Sparse Bayesian learning (SBL) integrates the Bayesian framework with the concepts of sparse representations and compressive sensing. SBL has been used for SSL \cite{gerstoft2016multisnapshot, nannuru2018sparse, ping2020three,xenaki2018sound}.
 However, the performance of these hybrid techniques remains unstable since the feature extraction routine varies across different datasets. 

Advancements in deep learning have led to the development of convolutional neural network (CNN) based methods for sound source localization. The CNN designed by \cite{chakrabarty2019multi} uses the multichannel STFT phase spectrograms to predict multi-speakers' azimuth in reverberant environments. The model consists of three convolutional layers with 64 filters of size ${2}\times{1}$ to consider neighboring frequency bands and microphones. Some deeper CNN architectures \cite{8668414, 9054754, 8683732, 8936994} are applied to estimate both the azimuth and elevation. Several three-dimensional convolutions networks \cite{9287344,9268154} report that networks for time, frequency, and channel can achieve better accuracy than 2D convolutions. Focusing on binaural audio-visual localization, Binaural Audio-Visual Network (BAVNet) \cite{wu2021binaural}, for pixel-level sound source localization using binaural recordings and videos, which significantly improves performance over traditional monaural audio methods, especially when visual information quality is limited. As a data-driven DL method, NI-CNN can learn latent features for azimuth prediction from human auditory nerve representations \cite{van2022goal}. These studies highlight the importance of advanced neural network architectures and feature extraction methods for enhancing the accuracy and resolution of sound source localization systems.

Transformer was initially proposed in natural language processing to handle long-range dependencies \cite{vaswani2017attention,devlin2018bert}. Recently, the Transformer was successfully applied in computer vision by casting images into patch embedding sequences \cite{dosovitskiy2020image,han2022survey}. Many hybrid models combined the Transformer with a CNN or Recurrent Neural Network (RNN) in audio processing, and some studies even directly embedded attention blocks into CNN or RNN to capture global features in a parameter-efficient way \cite{zhang2021attention,lin2020audiovisual,kong2020sound,9287224}. Transformer-based models in sound source localization have gained significant attention in recent years.\cite{yalta2021hitachi} uses a transformer encoder with residual connections and evaluates various configurations to manage multiple sound events. PILOT \cite{schymura2021pilot} is a transformer-based framework for sound event localization, capturing temporal dependencies through self-attention mechanisms and representing estimated positions as multivariate Gaussian variables to include uncertainty. Transformer-based models in sound event tasks have gained significant attention in recent years. The Audio Pyramid Transformer (APT) \cite{xin2022audio} with attention mechanism for weakly supervised sound event detection and audio classification, highlighting the application of transformer-based models in audio tasks. Multi-head self-attention, the parallel use of several attention layers in transformers, has also been used in SSL. Employing the first-order Ambisonic signals, Subsequently, the authors in \cite{gong2021ast} introduced AST which uses a Transformer model and variable-length monaural spectrograms to perform sound classification tasks. AST uses the overlapped-patch embedding generation policy to convert the intra-patch local features to inter-patch attention weights as a convolution-free, pure attention-based model. AST has achieved state-of-the-art \cite{piczak2015esc,warden2018speech} results on multiple datasets for audio classification tasks. 

The Vision Transformer (ViT) \cite{dosovitskiy2020image} represents a significant shift in the architecture of deep learning models for computer vision tasks. ViT divides an image into a sequence of fixed-size patches, linearly embedding them, and then processes them as tokens in a standard Transformer model. This method leverages the self-attention mechanism to capture long-range dependencies and contextual information across the image. The Audio Vision Transformer (AViT) is a model architecture that extends the concepts of Vision Transformers (ViTs) into the domain of audio processing.  The audio-spectrogram vision transformer (AS-ViT) \cite{ariyanti2023abnormal} use vision transformer models to analyze audio-spectrogram images for identifying abnormal respiratory sounds. The potential of ViT in audio-visual tasks such as sound source localization has been recognized \cite{lin2023vision}. Additionally, HTS-AT \cite{9746312}, a hierarchical token-semantic audio transformer, was designed to reduce the model size and training time, addressing the limitations of existing audio transformers. Binaural sound localization in noisy environments has been investigated using Frequency-Based Audio Vision Transformer (FAViT) \cite{phokhinanan2023binaural}. FAViT uses selective attention mechanisms inspired by the Duplex Theory, outperforming recent CNNs and standard audio ViT models in localizing noise speech. ViT-based localization has also been explored for through-ice or underwater acoustic tracking \cite{whitaker2022through}.

\section{Method}

\subsection{Model architecture}
\label{model}

The proposed Binaural Audio Spectrogram Transformer (BAST), is illustrated in Fig. \ref{fig:archi}. Similar to NI-CNN \cite{van2022goal}, a dual-input hierarchical architecture is utilized to simulate the human subcortical auditory pathway. As opposed to NI-CNN which uses convolution layers, here three Transformer Encoders (i.e., left, right and center), hereafter called TE-L, TE-R and TE-C are utilized to construct a pure attention-based model. In particular, the pre-processed left and right sound waves are converted to left and right spectrograms denoted by $x^{L} \in \mathbb{R}^{H \times T}$ and $x^{R}\in \mathbb{R}^{H \times T}$. Here, $H$ indicates the frequency band and $T$ indicates the number of Tukey windows ( with shape parameter: 0.25). 

In what follows, the TE-L path to process the input data is explained. The other path, i.e TE-R, follows the same process.
At the beginning of patch embedding layer, the left spectrogram $x^{L} \in \mathbb{R}^{H \times T}$ is first decomposed into an overlapped-patch sequence $x_{patch}^L \in \mathbb{R}^{P^2 \times (N_H N_T)}$, where $P$ is the patch size, $N_H$ and $N_T$ are the number of patches in height and width respectively obtained as follows, 

\begin{equation}
    \label{eq_patch_num}
    N_H = \left\lceil \frac{H-P+S}{S} \right\rceil ,
    N_T = \left\lceil \frac{T-P+S}{S} \right\rceil.
\end{equation}

In case $H-P$ and $T-P$ are not divisible by the stride $S$ between patches, the spectrogram is zero-padded on the top and right respectively. A trainable linear projection is added to flatten each patch to a $D$ dimensional latent representation, hereafter called patch embeddings \cite{dosovitskiy2020image}. Since our model outputs the sound location coordinates, the classification token in the Transformer encoder is removed. A fixed absolute position embedding \cite{dosovitskiy2020image} is added to the patch embeddings to capture the position information of the spectrogram in the Transformer. Here, the learnable position embedding is not used as it did not significantly change model performance compared to absolute position embedding\cite{gong2021ast}. The output of the left position embedding layer $z_{in}^L \in \mathbb{R}^{D \times (N_H N_T)}$ is then fed to the Transformer encoder TE-L. 


We use the identical Transformer encoder design in \cite{dosovitskiy2020image,gong2021ast}, consisting of $K$ stacked Multi-head Self-Attention (MSA) and Multi-Layer Perceptron (MLP) blocks. The BAST model performance is compared when using shared and non-shared parameters across the left and right Transformer encoders. Hereafter, BAST-SP refers to BAST model whose left and right Transformer encoders share parameters whereas in BAST-NSP the parameters of the left and right Transformer encoders are not shared.
The output of left and right Transformer encoder, i.e., $z_{out}^L$ and $z_{out}^R$, represents neural signals underlying the initial auditory processing stage along the left and right auditory pathway respectively. Subsequently, these binaural feature maps are integrated to simulate the function of the human olivary nucleus. Similar to NI-CNN, here three integration methods, i.e. addition, subtraction and concatenation, are investigated.  Specifically, addition is the summation of feature maps of both sides; subtraction represents left feature map subtracted from the right feature map; concatenation is implemented by concatenating $z_{out}^L$ and $z_{out}^R$ along the the first dimension to produce $z_{in}^C \in \mathbb{R}^{2D \times (N_H N_T)}$. TE-C receives the integrated feature map $z_{in}^C$ and output sequence $z_{out}^C$. Next, an average operation of the patch dimension and a linear transformer layer is applied to finally produce the sound location coordinates $(x, y)$. The last linear layer does not have any activation function, therefore the estimated coordinates can be any point on the 2D plane.

\subsection{Loss Function}
Three loss functions, i.e., Angular Distance (AD) loss \cite{xiao2015learning}, Mean Square Error (MSE) loss, as well as hybrid loss with a convex combination of AD and MSE, are explored in training the proposed model. Let $C_i=(x_i,y_i)$ and $\hat{C}_i=(\hat{x}_i,\hat{y}_i)$ denote the ground truth and predication coordinates for the i-$th$ sample. MSE loss measures the squared difference of Euclidean distance between the prediction and the ground truth as follows:
\begin{equation}
    \label{loss_mse}
    \textrm{MSE} = \frac{1}{N} \sum_i^N \|  C_i - \hat{C}_i \|_2^2,
\end{equation}
where $(\hat{x_i}, \hat{y_i})$ is the predicted coordinates, $(x_i, y_i)$ is the true sound location and $N$ is the batch size. Note that MSE loss is able to penalize the large Euclidean distance error but is insensitive to the angular distance, which means that the azimuth may differ with the same MSE. In contrast to MSE, AD loss merely measures the angular distance while ignoring the Euclidean distance:
\begin{equation}
    \label{loss_ad}
    \textrm{AD} = \frac{1}{\pi N} \sum^N_i \arccos{(\frac{C_i\hat{C}_i^T}{\|C_i\|_2 \|\hat{C}_i\|_2})}, 
\end{equation}
where $C_i,\hat{C}_i  \neq 0$. 


\begin{table*}
\centering
\caption{The performance of BAST-NSP and BAST-SP compared to NI-CNN and NI-CNN$^*$ when different loss and binaural integration methods are used. The best performed model in AD and MSE are in bold. The ↓ indicates the lower the value of the metric, the better the model performance.}
\label{table_performance}
\begin{tabular}{cccccccccc} 
\toprule
\multirow{2}{*}{\textbf{Model}}                                                  & \multirow{2}{*}{\textbf{Loss}} & \multicolumn{4}{c}{\textbf{Angular Distance(AD) ↓}} &  \multicolumn{4}{c}{\textbf{Mean Squared Error (MSE) ↓}}  \\ 
\cmidrule{3-10}
                                                                                 &                                &SS &Concatenation & Addition & Subtraction              &SS&Concatenation & Addition & Subtraction                   \\ 
\midrule
\multirow{3}{*}{\begin{tabular}[c]{@{}c@{}}CNN\cite{chakrabarty2019multi}$^*$\end{tabular}} & MSE                 &3.90°   & —     & —    & —                   &0.010  & —      & —    & —                         \\
                                                                                 & AD                             &42.69°   & —     & —    & —                   & —     & —      & —    & —                             \\
                                                                                 & Hybrid                         &3.09°    & —     & —    & —                   &0.011  & —      & —    & —                         \\ 
\midrule
\multirow{3}{*}{\begin{tabular}[c]{@{}c@{}}FAVit\cite{phokhinanan2023binaural}$^*$\end{tabular}} & MSE            &6.26°    & —     & —    & —                   &0.022  & —      & —    & —                         \\
                                                                                 & AD                             &17.37°   & —     & —    & —                   & —     & —      & —    & —                             \\
                                                                                 & Hybrid                         &3.73°    & —     & —    & —                   &0.015  & —    & —       & —                         \\ 
\midrule
\multirow{2}{*}{\begin{tabular}[c]{@{}c@{}}NI-CNN\cite{van2022goal}\end{tabular}}  & MSE                           &— & 4.80°         & 4.80°    & 5.30°                    & —& 0.011         & 0.013    & 0.014                         \\
                                                                                 & AD                             &— & 3.70°         & 3.90°    & 5.20°                    &— & —             & —        & —                             \\ 
\midrule
\multirow{3}{*}{\begin{tabular}[c]{@{}c@{}}NI-CNN\cite{van2022goal}$^*$\end{tabular}} & MSE                          & — & 8.92°         & 3.51°    & 3.67°                   & —& 0.077         & 0.032    & 0.038                         \\
                                                                                 & AD                             &— & 7.85°         & 1.97°    & 1.85°                  & — & —             & —        & —                             \\
                                                                                 & Hybrid                         &— & 8.35°         & 3.53°    & 3.19°                   &— & 0.074         & 0.033    & 0.031                         \\ 
\midrule
\multirow{3}{*}{BAST-NSP}                                                            & MSE                         &  — & 2.78°        & 2.48°    & 2.42°                  & — & 0.003         & 0.002    & 0.002                         \\
                                                                                 & AD                             & —& 2.39°        & 1.30°    & 1.63°                   &— & —             & —        & —                             \\
                                                                                 & Hybrid                        & —& 2.76°         & 1.83°    & \textbf{1.29°}          &— & 0.004         & 0.002    & \textbf{0.001}                \\ 
\midrule
\multirow{3}{*}{BAST-SP}                                                         & MSE                           & —& 2.02°         & 4.97°    & 1.94°                   & —& 0.002         & 0.018    & 0.002                         \\
                                                                                 & AD                            & —& 2.66°         & 13.87°   & 1.43°                   &— & —             & —        & —                             \\
                                                                                 & Hybrid                        &— & 1.98°         & 5.72°    & 2.03°                   &— & 0.003         & 0.026    & 0.002                         \\
\bottomrule
\end{tabular}
\flushleft{NI-CNN$^*$ uses spectrogram as model input.}
\flushleft{SS represents singular stream.}
\end{table*}

\begin{table}
\centering
\caption{The number of layers in each Transform encoder as well as the total number of trainable parameters of the proposed models. Tuple ($\cdot$, $\cdot$, $\cdot$) indicates the number of layers in the left, right and center Transformer encoder respectively.}
\label{table_prams}
\begin{tabular}{cccc} 
\toprule
\textbf{Model}       & \begin{tabular}[c]{@{}c@{}}\textbf{Interaural}\\\textbf{Integration}\end{tabular} & \begin{tabular}[c]{@{}c@{}}\textbf{Transformer}\\\textbf{ Layers }\end{tabular} & \begin{tabular}[c]{@{}c@{}}\textbf{Trainable}\\\textbf{Parameters}\end{tabular}  \\ 
\midrule
\multirow{3}{*}{BAST-NSP}                 & Concatenation                                                                                      & (3, 3, 3)                                                                       & \textasciitilde{}76M                                                           \\
\multicolumn{1}{l}{} & \begin{tabular}[c]{@{}c@{}}Addition/\\Subtraction\end{tabular}                                     & (3, 3, 3)                                                                       & \textasciitilde{}57M                                                           \\ 
\midrule
\multirow{3}{*}{BAST-SP}              & Concatenation                                                                                      & (3, 3, 3)                                                                       & \textasciitilde{}57M                                                           \\
\multicolumn{1}{l}{} & \begin{tabular}[c]{@{}c@{}}Addition/\\Subtraction\end{tabular}                                     & (3, 3, 3)                                                                       & \textasciitilde{}38M                                                           \\
\bottomrule
\end{tabular}
\end{table}

\section{Experiments}

\subsection{Dataset}
We use the binaural audio data in \cite{van2022goal}, which consists of a training dataset and an independent testing dataset. In the training dataset, 4600 real-life sound waves (duration:500 ms, sampling rate: 16000) are placed in 36 azimuth positions, respectively, with $10\degree$ azimuth resolution, $0\degree$ elevation, and 1-meter distance from the center point. In addition, sound waves are spatialized with two acoustic environments, i.e. an anechoic environment (AE) without reverberation and a 10m $\times$ 14m lecture hall with reverberation (RV). In particular, here the training and test sets contain data from both AE and RV environments. In total, the training dataset has 331200 binaural learning samples. Similarly, the independent testing dataset contains 400 new sound waves processed with the same method as described above, producing 28800 testing samples.

\subsection{Baseline methods}
In this study, we establish a comprehensive framework for evaluating the performance of our proposed model by comparing it against four baseline models widely utilized in the field. Four baseline models were employed in this work: two-stream CNN-based models NI-CNN$^{*}$ \cite{van2022goal} and NI-CNN \cite{van2022goal}, one-stream CNN model \cite{chakrabarty2019multi}, and ViT-based FAViT \cite{phokhinanan2023binaural}. NI-CNN and NI-CNN$^{*}$ models use cochleogram and spectrogram as model inputs, respectively. 
The CNN and FAVit model inputs are spectrogram. The hyper-parameters of all baseline models are empirically found to be optimized. By benchmarking our proposed model against these established baselines, we provide a comprehensive evaluation framework to assess its efficacy.

\subsection{Model Evaluation}
Models were evaluated by means of MSE and AD errors defined in Eq. (\ref{loss_mse}) and (\ref{loss_ad}). The lower AD and MSE errors the better localization performance is. Note that the MSE metric has no meaning when BAST is trained by AD loss because this loss does not optimize the Euclidean distance between the ground truth and the prediction, and that BAST has no constraints on the numerical range of the predicted coordinates. The one-stream models CNN and FAViT have no concatenation, addition, or subtraction modes, so the MSE and AD of these two models are measured only once.

\subsection{Training Settings}
\label{settings}
As mentioned in \ref{model}, each sound wave are transformed to binaural spectrogram (size: 2$\times$129$\times$61, frequency range: 0-8000Hz, window length: 128ms, overlap: 64ms) before training. It is important to note that while the STFT transformation employed may reduce the fine-grained temporal differences between channels, our focus predominantly lies on interaural level differences (ILDs) rather than interaural time differences (ITDs). In order to have balanced training samples, we randomly select 75\% binaural spectrograms in each azimuth position and listening environments of training dataset. The remaining data is used as validation set. As stated before in the Dataset section, a separate test dataset is available for this study. This setting results in $248400$ training samples and $82800$ validation samples. 
Here, Adam optimizer \cite{kingma2014adam} is used to train the model for 50 epochs with a batch size of 48 and a fixed learning rate of 1e-4. In patch embedding layers, the stride of patches is set to 6, yielding 180 patches per spectrogram. Each Transformer encoder has three layers, with 1024 hidden dimensions (2048 dimensions when using concatenation as integration method in the last Transformer encoder TE-C), 16 attention heads in MSA blocks, 1024 dimensions in MLP blocks, 0.2 dropout rate in patch embeddings and MLP blocks. Our model implementation \footnote{code available at \href{https://github.com/ShengKuangCN/BAST}{https://github.com/ShengKuangCN/BAST} } is based on Python 3.8 and Pytorch 1.9.0, and are trained from scratch on 2 $\times$ NVIDIA GeForce GTX 1080Ti GPUs with 11GB of RAM.  The empirically found and used number of layers in each Transformer encoder as well as the total number of trainable parameters are presented in Table \ref{table_prams}.

\begin{figure}
    \centering
    \includegraphics[scale=0.48]{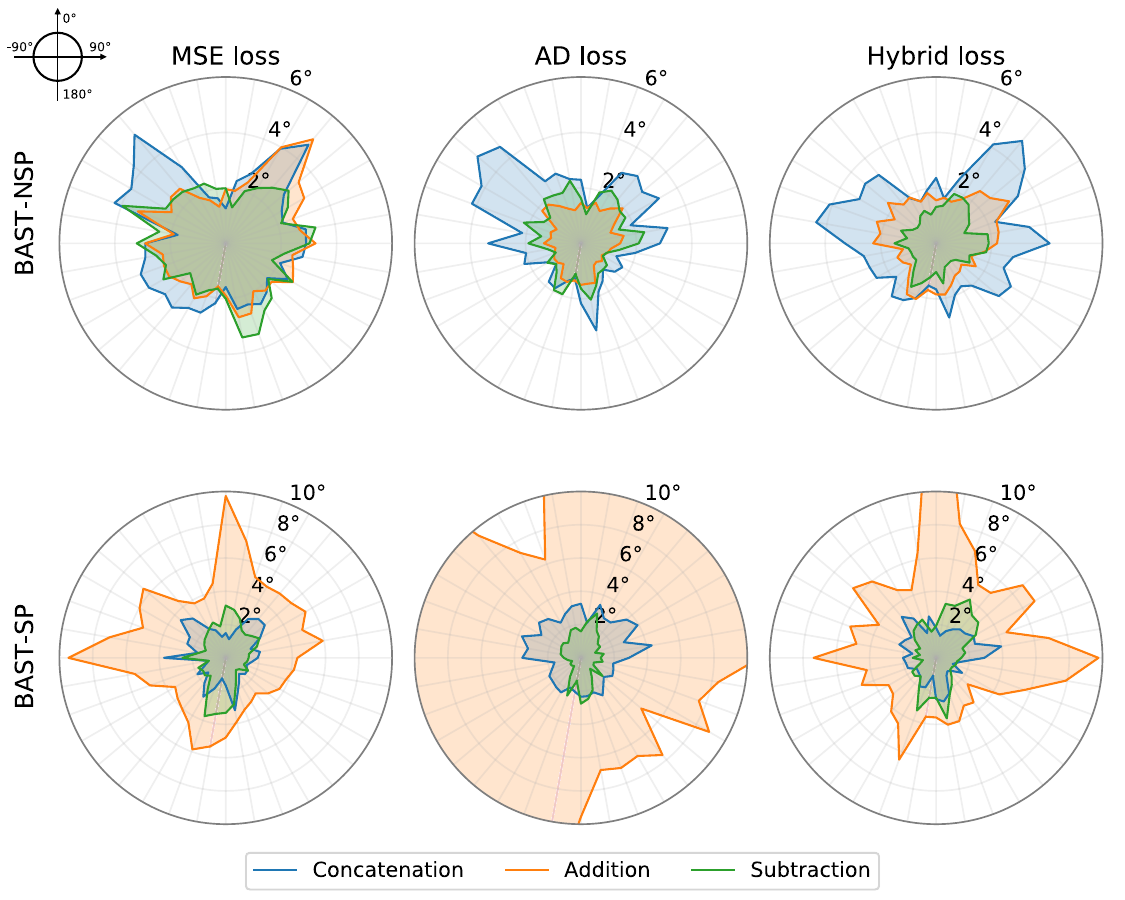}
    \caption{The angular distance (AD) error of the proposed BAST in each azimuth with different loss functions and interaural integration methods.}
    \label{fig:radar_plot}
\end{figure}

\begin{figure}[!h]
    \centering
    \includegraphics[scale=0.45]{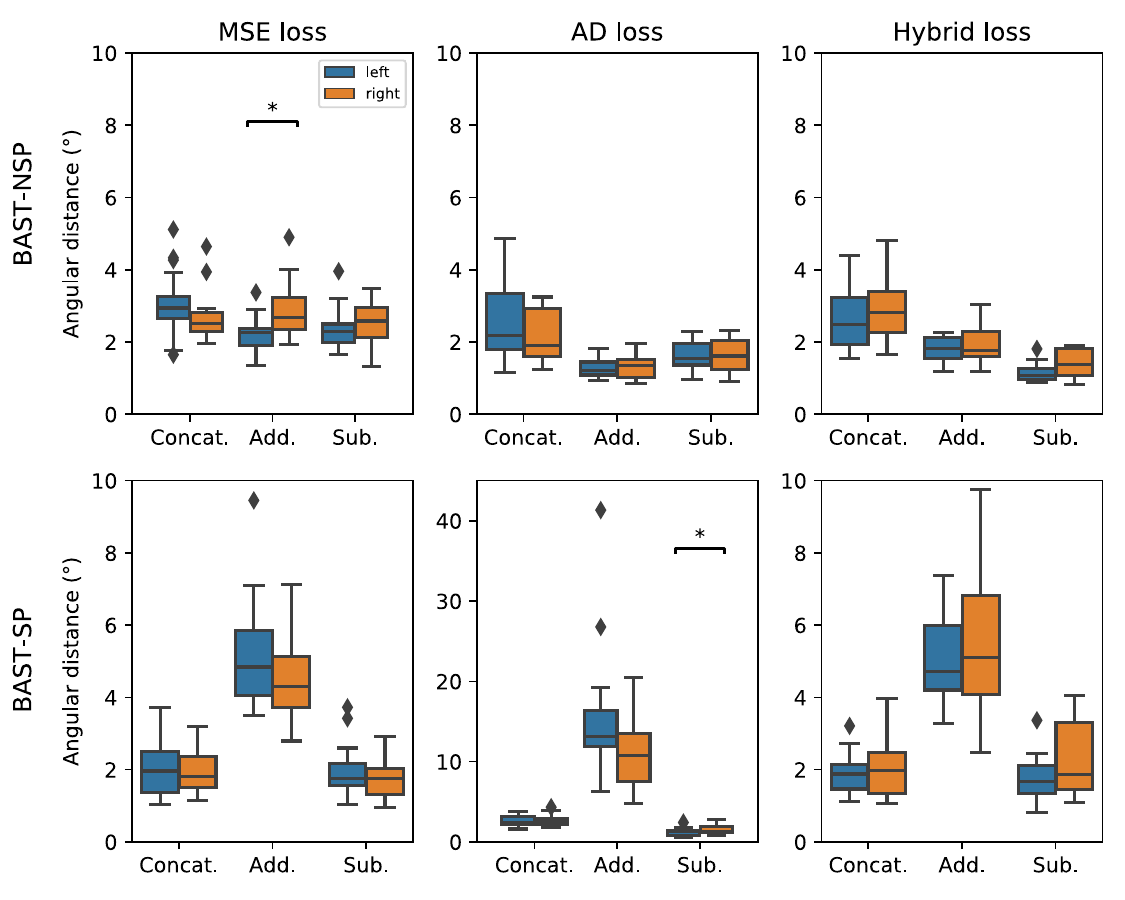}
    \caption{The AD error of the proposed BAST-NSP and BAST-SP in the left and right hemifield. The boxplot indicates quartiles of the metric distribution with respect to azimuths. The asterisk between two boxes indicates the statistical significance (p$<$0.05, paired t-test with FDR correction) between the left and right hemifield.}
    \label{fig:left_right_ad}
\end{figure}

\begin{figure}[!t]
    \centering
    \includegraphics[scale=0.45]{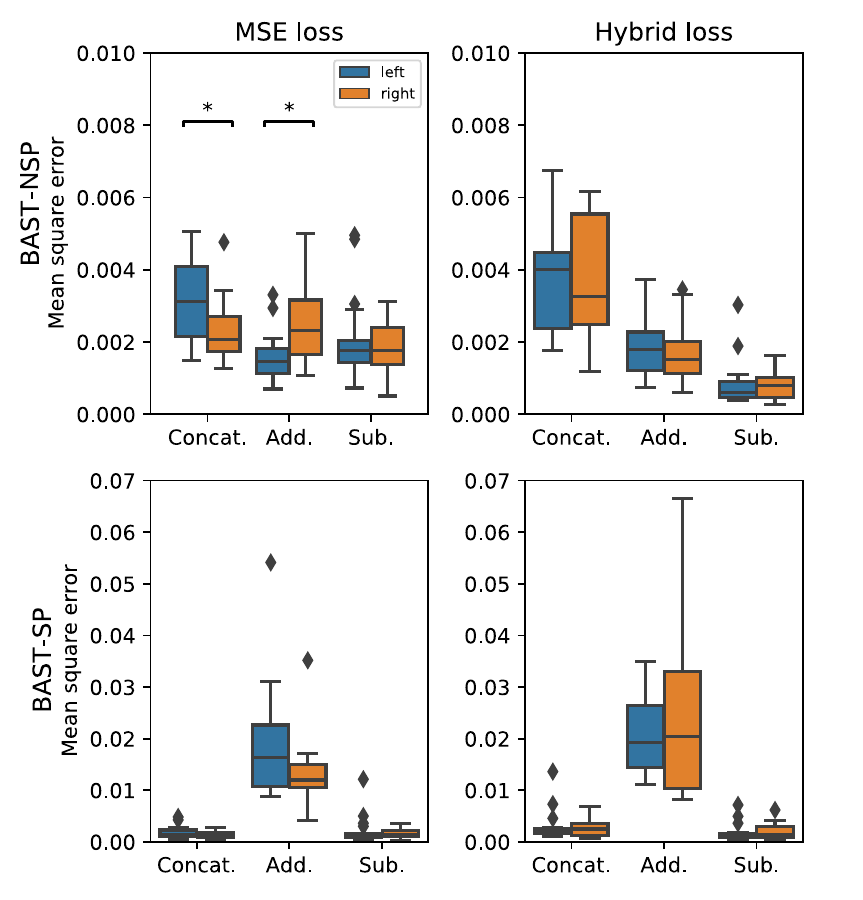}
    \caption{The MSE of the proposed BAST-NSP and BAST-SP in the left and right hemifield. The boxplot indicates quartiles of the metric distribution with respect to azimuths. The asterisk between two boxes indicates the statistical significance (p$<$0.05, paired t-test with FDR correction) between the left and right hemifield. 
    }
    \label{fig:left_right_mse}
\end{figure}

\begin{figure*}[!h]
    \centering
    \includegraphics[scale=0.8]{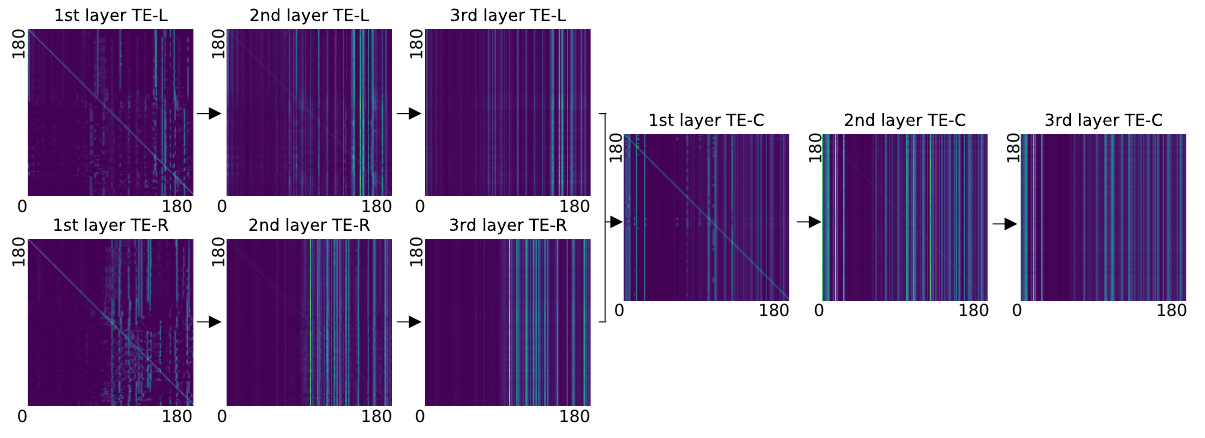}
    \caption{An example of the attention matrices in the proposed model (i.e., BAST-NSP, hybrid loss and subtraction). The corresponding sound clip was randomly selected in the category of human speech with reverberation. For each layer, we present the patch-to-patch attention matrix (size: 180$\times$180) calculated by the rollout method in \cite{chefer2021transformer}. Note that we initialize the attention matrix at the first layer of TE-C by summing the attention matrices at the last layer of TE-L and TE-R.}
    \label{fig:attn_flow}
\end{figure*}

\section{Results}

\subsection{Overall Performance}

The proposed BAST-NSP and BAST-SP models' performance is compared with those of the NI-CNN, CNN, and FAVit models. In particular, for the NI-CNN model, two modes of implementations corresponding to correlogram and spectrogram as model inputs have been considered and are denoted by NI-CNN and NI-CNN$^*$, respectively. The obtained results of the compared models with different combinations of binaural integration methods and loss functions are tabulated in Table \ref{table_performance}. BAST-NSP has achieved the best AD error of 1.29\degree and the best MSE of 0.001 when using the subtraction binaural integration and the hybrid loss function. Compared to NI-CNN, BAST-NSP reduces AD error 65.4\% from 3.70\degree to 1.29\degree and MSE 90.9\% from 0.011 to 0.001. In addition, BAST-NSP outperforms NI-CNN$^*$ (NI-CNN$^*$: AD=1.85\degree, MSE=0.031), although both models received the same input. BAST-SP has achieved AD=1.43\degree, MSE=0.002, surpassing the performance of NI-CNN and NI-CNN$^*$ models while still inferior to BAST-NSP. In this study, BAST-NSP outperforms the other tested models in performing binaural sound localization.

Two-stream models (NI-CNN, NI-CNN$^*$, BAST-NSP, BAST-SP) outperform one-stream models (CNN, FAVit) in both Angular Distance (AD) and Mean Squared Error (MSE). BAST-NSP shows the best overall performance among the two-stream models, followed closely by BAST-SP. BAST-NSP and BAST-SP show significant improvements in AD, especially in the hybrid loss function, with BAST-NSP's best AD at 1.29\degree and BAST-SP's best AD at 1.43\degree, compared to the best one-stream AD of 3.09\degree (CNN). Two-stream models generally have lower MSE compared to one-stream models. BAST-NSP has the lowest MSE, with the best performance at 0.001 (Hybrid loss with Subtraction), compared to the lowest MSE of one-stream models at 0.010 (CNN). One-stream models have shown more poor performance in AD loss than the best-performing two-stream models.

We further analyze the influence of different binaural integration methods on the BAST-NSP and BAST-SP performance. Here, the performances are compared in terms of AD error. Specifically, in both cases when the BAST-NSP is trained by AD loss and hybrid loss, binaural integration through addition and subtraction improved the model performance compared to concatenation (AD loss: Add.=1.30\degree, Sub.=1.63\degree, Concat.=2.39\degree; Hybrid loss: Add.=1.83\degree, Sub. =1.29\degree, Concat.=2.76\degree, see Table \ref{table_performance}). In case of MSE loss, the performance across the three integration methods of BAST-NSP is similar. In BAST-SP, addition integration causes a huge AD error increment over BAST-NSP (BAST-SP: 4.97\degree, BAST-NSP: 1.30\degree), indicating that the left-right identical feature addition brings a great challenge to the model to predict the azimuth.
    
The effect of three types of loss functions on BAST-NSP and BAST-SP performance are not the same. In BAST-NSP, AD loss achieves lower AD when using concatenation or addition, but the hybrid loss yields the lowest AD in terms of subtraction. In BAST-SP, one can observe an interaction of loss function and binaural integration methods, i.e. the best loss function depends on the applied binaural integration methods.

\subsection{Performance at different azimuths}
To better understand the localization performance of the models in different azimuth, the test AD error of each azimuth is shown in Fig. \ref{fig:radar_plot}. The test AD error in BAST-NSP is much smaller when the sound source is located closer to the interaural midline. This the error pattern for BAST-NSP is similar as for humans, highlighting the relevance of independent processing in the left and right stream. However, this pattern is not observed in BAST-SP. 

\begin{figure}
    \centering
    \includegraphics[scale=0.5]{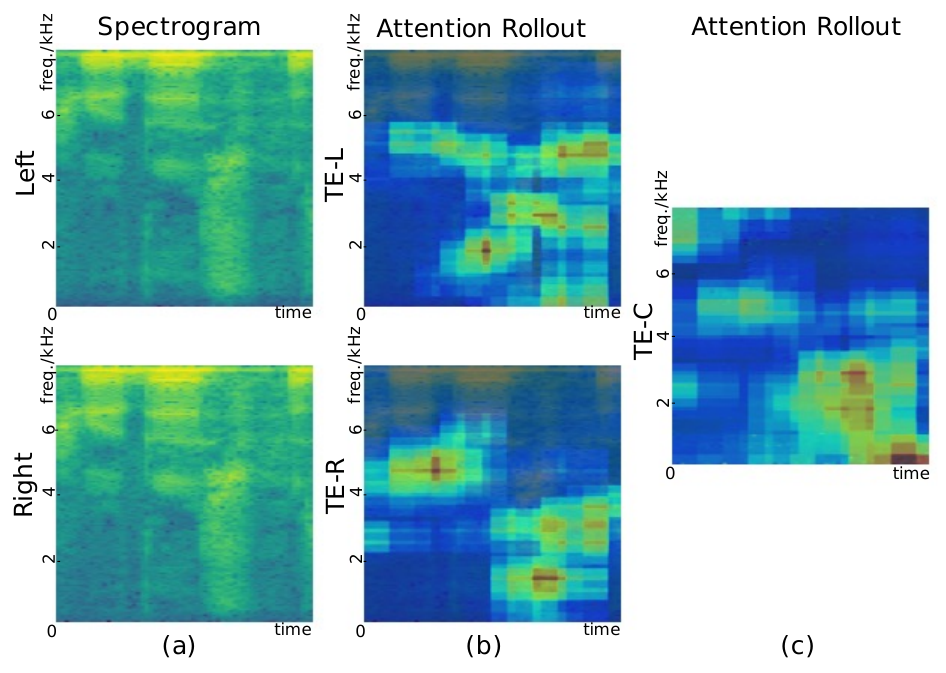}
    \caption{Attention rollout corresponding to the spectrogram shown in Fig. \ref{fig:attn_flow}. (a): Left and right spectrogram. (b): The left and right attention rollout obtained from the 3rd layer of TE-L and TE-R Transformers. (c): The left and right Attention rollout obtained from the 3rd layer of TE-C.}
    
    \label{fig:attn_rollout}
\end{figure}

\subsection{Performance in left and right hemifield}
To explore the symmetry of the model predictions, we further compare the evaluation metrics between the left-right hemifields. One can observe comparable model performance in left and right hemifield in Fig. \ref{fig:left_right_ad} and \ref{fig:left_right_mse}. This result is confirmed by paired t-test (False Discovery Rate (FDR) corrected for multiple comparisons). More specifically, Fig. \ref{fig:left_right_ad}, shows an insignificant difference of AD error between the left and right hemifield in most conditions (corrected p$>$0.05). However, a minor but significant difference (corrected p$<$0.05) is observed in BAST-NSP trained with MSE loss and addition integration, and in BAST-SP trained with AD loss and subtraction. 
More precisely, the difference in AD error between the left and right hemifield was not significant in most conditions (corrected p$>$0.05, Fig. \ref{fig:left_right_ad}), thus supporting the consistent symmetry of model predictions.

\subsection{Performance in different environments}
We conduct two additional experiments to illustrate the generalization of the proposed model by training in one listening environment and testing in both environments separately, i.e., AE and RV. This analysis is conducted on the best performing model, i.e., BAST-NSP with hybrid loss and subtraction integration method. 
As shown in Table \ref{table_env}, the model that is trained using the data of both AE and RV environments, achieves the best test results compared to other models which are trained only using the data of one of the environment. 

\begin{table}
\centering
\caption{The performance of the proposed BAST-NSP model in different listening environments. AE and RV indicate the anechoic and reverberation environments respectively. 
}
\label{table_env}
\begin{tabular}{cccc} 
\toprule
\begin{tabular}[c]{@{}c@{}}\textbf{Training}\\\textbf{ Environment}\end{tabular} & \begin{tabular}[c]{@{}c@{}}\textbf{Testing}\\\textbf{ Environment}\end{tabular} & \textbf{AD} & \textbf{MSE}  \\ 

\midrule
\multirow{2}{*}{AE}                                                              & AE                                                                              & 1.14°       & 0.001         \\
                                                                                 & RV                                                                              & 8.66°       & 0.027         \\ 
\midrule
\multirow{2}{*}{RV}                                                              & AE                                                                              & 16.70°      & 0.078         \\
                                                                                 & RV                                                                              & 1.65°       & 0.002         \\ 
\midrule
\multirow{2}{*}{AE+RV}                                                           & AE                                                                              & \textbf{1.10°}       & \textbf{0.001}         \\
                                                                                 & RV                                                                              & \textbf{1.48°}       & \textbf{0.001}         \\

\bottomrule
\end{tabular}
\end{table}

\subsection{Attention Analysis}
To interpret the localization process, we utilize Attention Rollout \cite{abnar2020quantifying} to visualize the attention maps of the proposed model (BAST-NSP with subtraction method and hybrid loss). Rollout calculates the attention matrix by recursively multiplying the attention matrices along the forward propagation path. \cite{chefer2021transformer} enhanced this method by adding an additional identical matrix before multiplication to simulate the effect of residual connection of MSA. Due to the interaural integration layer, in BAST-NSP, we initialize the attention matrix of TE-C by summing the attention weights from both sides regardless of the integration method.

Fig. \ref{fig:attn_flow} shows the patch-to-patch attention matrices (size: 180$\times$180) of a randomly selected spectrogram. In the first layer of each Transformer, most of the patches are self-focused and pay attention to some scattered patches. However, in the last layer, all patches yield nearly consistent attention weights to some specific patches. The attention rollout heat map with respect to the left and right spectrogram are depicted in Fig. \ref{fig:attn_rollout}. Although parameter-sharing setting is not used in BAST-NSP, one can still observe that the model focuses most of its attention on similar regions on both sides, see Fig. \ref{fig:attn_rollout}(b). The final attention map, Fig. \ref{fig:attn_rollout}(c), shows that the model further processes the attention after the integration layer and boosts the attention weights in bottom left regions.

\section{Conclusion}
In this paper, a novel Binaural Audio Spectrogram Transformer (BAST) for sound source localization is proposed. The obtained results show that this pure attention-based model leads to significant azimuth acuity improvement compared to CNN, FAVit and NI-CNN models. In particular, subtraction interaural integration and hybrid loss is the best training combination for BAST. Additionally, we found that the performance and statistical significance in left-right hemifields vary with different combinations of training settings. In conclusion, this work contributes to a convolution-free model of real-life sound localization. The data and implementation of our BAST model are available at
\url{https://github.com/ShengKuangCN/BAST}.



\bibliography{Main}

\end{document}